\documentclass[aps,showpacs,preprintnumbers,preprint]{revtex4}

\usepackage{graphicx}

\begin{document}

\title{Community structures and role detection in music networks}

\author
{T. Teitelbaum$^1$, P. Balenzuela$^1$, P. Cano$^{2,3}$ and Javier M. Buld\'{u}$^4$}

\affiliation{$^1$ Departamento de F\'{\i}sica, Facultad de Ciencias Exactas y Naturales, Universidad de Buenos Aires and CONICET, Buenos Aires, Argentina}
\affiliation{$^2$ Music Technology Group, Universitat Pompeu Fabra, Barcelona, Spain}
\affiliation{$^3$ BMAT, Barcelona Music and Audio Technologies, 08018 Llacuna 162, Barcelona, Spain}
\affiliation{$^4$ Complex Systems Group,
Universidad Rey Juan Carlos, Tulip\'an s/n,
28933 M\'ostoles, Madrid, Spain.}

\date{\today}

\begin{abstract}

We analyze the existence of community structures in two different social networks obtained 
from similarity and collaborative features between musical artists. Our analysis reveals some characteristic organizational patterns  and provides information about  the driving  forces behind the growth of the networks. In the similarity network, we find a strong correlation between clusters of artists and musical genres. On the other hand, the collaboration network shows two different kinds of communities: rather small structures related to music bands and geographic zones, and much bigger communities built upon collaborative clusters with a high number of participants related through the period the artists were active.
Finally, we  detect the leading artists inside their corresponding communities and analyze their roles in the network by looking at a few topological properties of the nodes.

\end{abstract}

\maketitle


{\bf

Music is one of the richest sources of interaction between individuals.
\textit{Besides the usual connections between artists and listeners, it is possible to have artist-artist and listener-listener relations}. In the current work we analyze artist-artist interactions and their implications in music
similarity and collaboration. To that end, we construct two different networks where nodes represent musical artists: the similarity network, where
artists are linked if a certain similarity exists between them (evaluated by musical editors) and the collaboration network, where a link exists between two artists if they have ever performed together.
We detect and analyze the internal communities that spontaneously arise in both networks, 
which are driven by musical/social ``forces", and  show that the appearance of these communities is strongly related to the existence of musical genres. Furthermore, we are able to discriminate the main actors in the formed structures and extract their role in the network through the calculation and classification of a few topological properties of the nodes.

}

\section{Introduction}

Since the seminal paper of Milgram \cite{mil67} investigating the flow of information through acquaintance networks,
social (complex) networks have attracted the interest of scientists in a variety of fields \cite{wat04}.
Many kinds of social structures arise when analysing the different types of interdependency among individuals (or organizations), such as financial exchange, friendship, kinship, sexual relations or disease transmission. 
In the current work we focus on those social networks where music is the driving force that generates interaction between individuals.
Specifically, we consider musical artists as the fundamental nodes of the network and a certain musical relation as the 
linking rule. Two different types of networks are obtained: first, the similarity network, where artists are linked if their music are somewhat
similar, and second, the collaboration network, where artists are linked if they have ever performed together. 
The relevance of these kinds of networks does not only rely on a social science perspective but also in musical aspects, such as
the understanding of musical genres \cite{lam05,lam06} or music recommendation \cite{can06}.  

\begin{figure}[t]
\includegraphics[width=130mm]{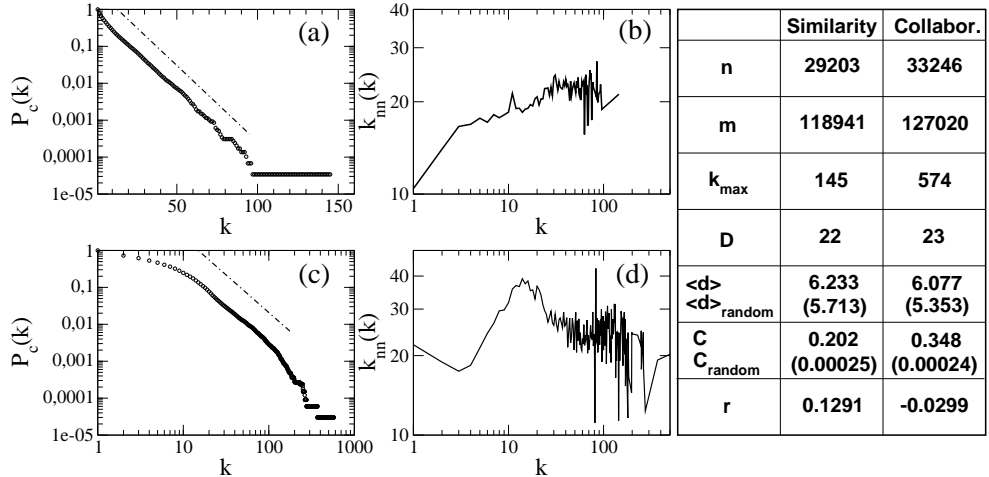}
\caption{(a) and (c) show the cumulative degree distribution $P_c(k)$ of
the similarity and collaboration networks respectively (note the different scale in the X-axis). (b) and (d) are their corresponding nearest neighbor degree distributions $k_{nn}(k)$.
Parameters shown in the table are: number of nodes (n), number of links (m), highest degree ($k_{max}$), diameter of the network (D), mean shortest path
($<d>$), clustering coefficient (C) and  Pearson correlation coefficient (r) \cite{new02b}.}
\label{fig:f01}
\end{figure}

Networks are obtained from the All-Music database of music metadata \cite{all}. The content of the database is created by 
professional editors and writers. Despite the linking rule being clear when creating the collaboration network, the similarity between artists is a more complex task.  A great deal of research is devoted towards the development of audio content-based algorithms capable of quantifying similarity between musical pieces \cite{foo97,blu99,log01}. 
Although great advances have been made in this field, the criterion
of musical experts still prevails over similarity software. If we translate the problem from musical pieces to musical artists \cite{ell02}, the evaluation of musical similarity becomes a subjective task where expert musical editors have the last say.


The intersection between both networks has been recently analyzed \cite{par07} from a complex network perspective \cite{new02,boc06}.
In the current work we go one step further by studying the structures that arise in the spontaneous 
organization of these particular social networks. Specifically, we are interested in the existence and characterization of communities
inside the network and the driving forces that induce their appearance. We also see how different kinds of community structures arise
at different partition levels and how they are related to the existence of musical genres (in the case of the similarity network) and
inter/intra band collaboration (in the case of the collaboration network).  
Figure \ref{fig:f01} summarizes the main  parameters of the network together with the cumulative degree distributions $P_c(k)$ and 
the nearest neighbor degree distributions $k_{nn}(k)$. Despite both networks sharing a small world topology \cite{str98}, there exist differences in their degree distribution and assortativity \cite{par07,new02b}. 

\section{Community detection}

\begin{figure}[!b]
\vskip -0cm
\includegraphics[width=85mm]{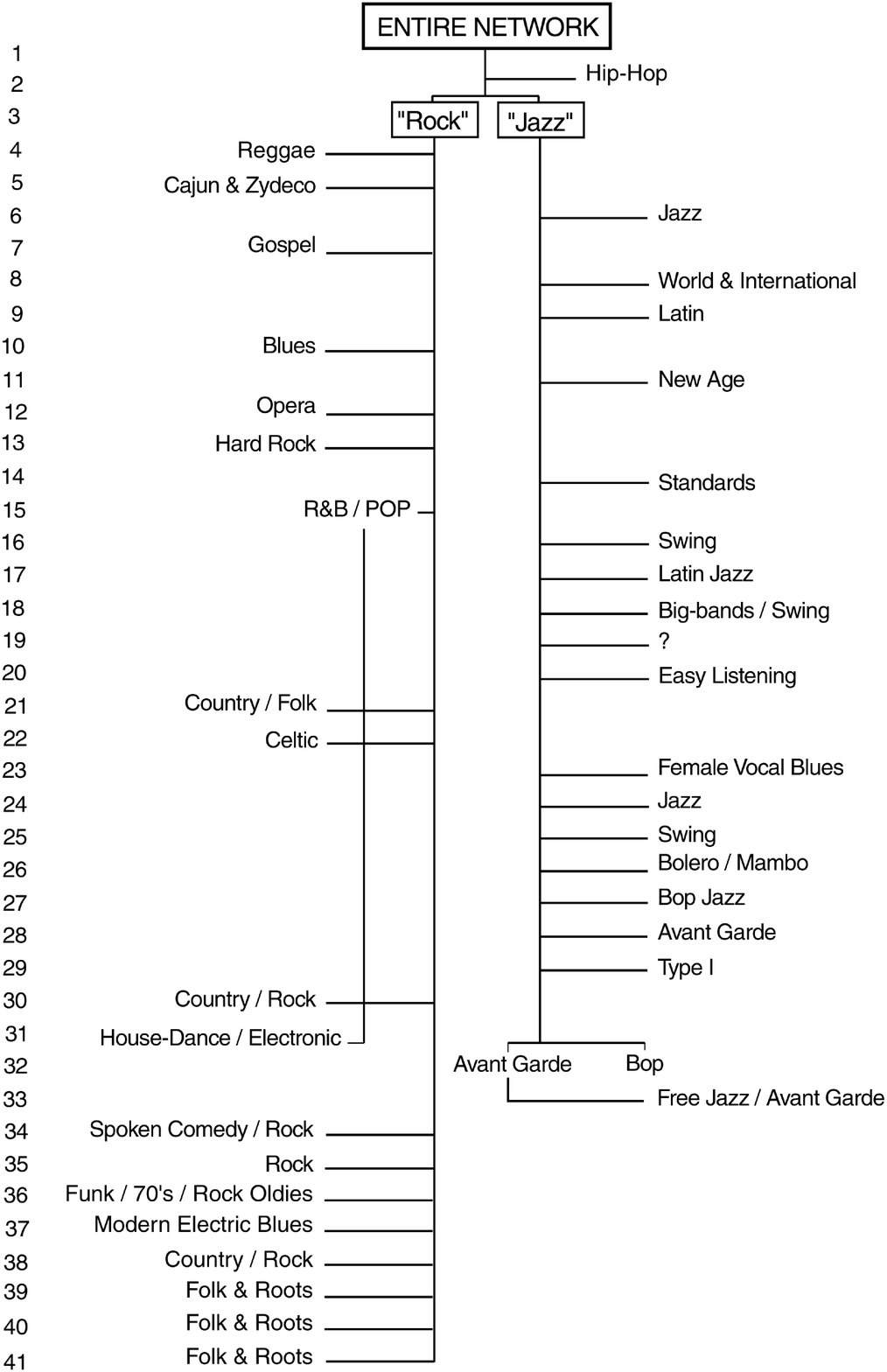}
\caption{Dendogram of communities detected in the similarity network when applying the GN algorithm. In every step (left column) a cluster (community) splits out from the network.}
\label{fig:dendo}
\end{figure}

Detection of communities in complex networks has gained a lot of attention 
during recent years \cite{gir02,pal05,dan05}, a fact  reflected in the existence
of several community-detection algorithms. Among them, we have selected the Girvan-Newman (GN) algorithm \cite{gir02} for its agreement between effectiveness and time consumption. As we will explain later, the GN is valid only for low to moderate values of the inter-community connections, which is the case of the networks analyzed here.

The GN algorithm is based on the sequential removal of those links
with the highest betweenness, which is measured as proportional to the number of shortest paths running along each link \cite{new04}. This way, the network breaks into isolated clusters (communities) which, in turn, can be further split in successive steps. 
In  Figure \ref{fig:dendo}, we plot this evolution for the similarity network. In order to understand the emergent communities, we use the fact that the All Music database tags each artist as belonging to one or more genres and we choose the most frequent tag to label each community.    We can identify the first split as a hip-hop community, followed by the division into two main groups dominated by ``rock" and ``jazz" artists respectively. In subsequent divisions there appear genres such as Blues, Opera or Hard Rock from the former ``rock" community, and Jazz, Latin-Bolero and Standards from the Jazz community.
 
In order to quantify the quality of the divisions we compute the modularity $Q$ of each partition. As explained in \cite{new04}, a modularity $Q=0$ indicates that the detected community structure is similar to the one existing in an equivalent random network or, in other words, links between nodes are randomly distributed and they are not related to the existence of certain cliques inside the network.  On the contrary, values approaching $Q=1$, which is the maximum, indicate strong community structure.

\begin{figure}[!b]
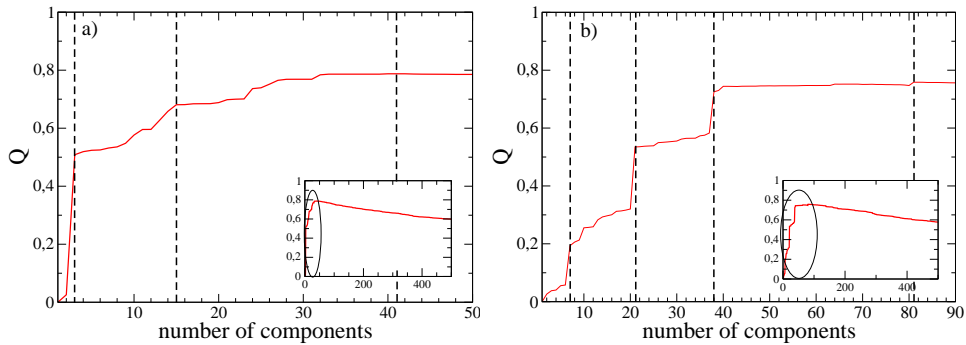

\centering
\includegraphics[width=63mm]{fig02a.eps}
\includegraphics[width=63mm]{fig02b.eps}
\caption{Modularity $Q$ of the communities (insets) as the GN algorithm splits the similarity (a) and collaboration (b) networks.
In the main plots, we have zoomed in on the region indicated in the insets, which correspond to the maximum of the $Q$ evolution. Dashed lines
indicate sudden increments of Q.}
\label{fig:f02}
\end{figure}

Figure \ref{fig:f02} shows the evolution of the modularity (Q) \cite{new04} as both networks are divided into independent clusters (by removal of links with the highest betweennes). 
We can observe the existence of sudden increments of Q  related to different satisfactory network partitions. As reported in \cite{new04} the absolute maximum is not always associated with the best partition, and therefore, each of these large jumps in $Q$ must be analyzed independently with regard to the nature of the data. 

As we saw in the dendogram of  the similarity network (Figure \ref{fig:dendo}), the possible partitions are related to the genre classification of the artists belonging to each detected community. The maximum value of $Q$ ($Q \simeq 0.79$) appears when the network is split into $41$ communities, all of them related to musical genres or styles within those genres.  However, the most significant partition is observed when the network is divided into  $15$ communities ($Q \simeq 0.68$) since each community can easily be described by a well defined musical genre. Further divisions of this network are mainly dominated by the appearance of different styles inside each genre.

In the case of the collaboration network the maximum appears for $81$ communities with a $Q=0.76$. In this case, the interpretation of the existing communities is more complex since several factors such as generational overlapping, geographical proximity, genre affinity, or, simply, the existence of music bands, induce community formation.

It is worth mentioning that the obtained values of modularity reveal a strong community structure. In all the mentioned cases the percentage of 
inter-communities links were always less than $17\%$. If we compare with toy-networks used to evaluate community detection algorithms
\cite{dan05}, we see that these values of inter-community links correspond to the region where the GN algorithm is as good as the others. This conclusion is also supported if we look at the inset of Figure 3 in \cite{arenas05}, where the authors show that values of modularity $Q$ greater than $0.5$ 
correspond to a region where the GN algorithm performs accurately.
All this evidence supports the use of the GN algorithm as a suitable community detector in these kinds of networks.

\begin{figure}[!b]
\vskip -0cm
\includegraphics[width=145mm]{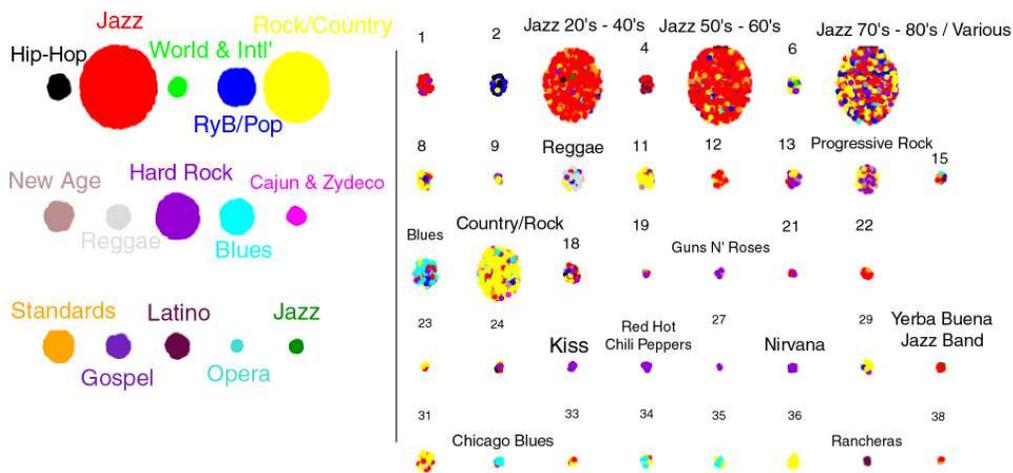}
\caption{Detected communities for the similarity (left) and collaboration (right) networks. Colors, which correspond to different musical
genres (similarity communities), are introduced to help comparison between similarity and collaboration communities.}
\label{fig:f03}
\end{figure}

In Fig. \ref{fig:f03} we plot the most significant partitions detected by the community structure algorithm, i.e., a division into $15$ communities
in the case of the similarity network (left plot) and $41$ communities in the collaboration network (right plot). Since each cluster
of the similarity network is related to a certain musical genre, we assign different colors to each community and we keep them in the collaboration
clusters. This way, we can observe how musical genres spread among the collaboration clusters and we can compare the relation between genres and collaborations.
Concerning the collaboration network, two kinds of communities are detected, one with
a small number of nodes corresponding to the existence of bands and geographic zones, and the other related to certain collaboration communities, where jazz artists are the most interactive nodes.
It is remarkable that two of the largest collaboration communities (3 and 5) are mainly formed by jazz players, a community of artists that presents a high degree of collaboration. We identify two kinds of ``collaborators" in these big communities,  one related to artists which usually play in several bands during their career (e.g., John Coltrane or Stan Getz) and the other related to jazz artists that usually perform as sessionist given that they are experts in one particular instrument (e.g., Paulinho Da Costa or Ron Carter).
Furthermore, these two largest communities correspond to different generations of jazz players, community 3 to the 20's-30's-40's and community 5 to the 50's-60's.
Interestingly, the community of jazz artists who performed together between 1912 and 1940 (which would correspond to community 3 of Fig.\ref{fig:f03})
was previously studied in \cite{gle03}.

\section{Role classification}

Once the existing communities have been identified we will try to
infer the artists' roles inside their communities by mere inspection of the network topology. Recently, Guimer\`a {\em et al.} \cite{gui05} have 
introduced a classification of the node 
functionality by analyzing the connectivity of nodes within the community structure.
Two properties of the node connectivity based on the inter/intra community connections are checked. One is the  
within-module degree $z_i$, which accounts for the connections of the node inside its community, and is defined as:
\begin{equation}
z_i=\frac{\kappa_i-\bar{\kappa}_{s_i}}{\sigma_{\kappa_{s_i}}}
\end{equation}
where $\kappa_i$ is the degree of node $i$, $\bar{\kappa}_{s_i}$ is the mean degree inside the community $s_i$ of node $i$
and $\sigma_{\kappa_{s_i}}$ is the standard deviation of $k$ in $s_i$.
High values of $z_i$ reflect that node $i$ is a well connected node inside its community (i.e., a hub), while negative values
indicate a connectivity below the average (peripheral nodes).

Another characteristic to be evaluated is how the links of a certain node are distributed between the communities. This is measured using the participation coefficient $P_i$ and accounts for the inter-community link distribution of node $i$:
\begin{equation}
P_i=1-\sum^{N_M}_{s=1} \left( \frac{\kappa_{is}}{\kappa_i} \right)^2
\end{equation}
where $N_M$ is the total number of communities, $\kappa_{is}$ is the number of links of node $i$ that are connected to nodes in community $s$ and 
$\kappa_i$ is the total degree of node $i$. The participation coefficient ranges from zero (all links inside its own community) to close to unity (all links equally distributed among all communities).

In the role classification proposed by Guimer\`a {\em et al.} the functionality is obtained by analyzing the position of nodes in a two dimensional space given by
$(P_i,z_i)$. Nodes with $z \geq 2.5$ are considered hubs and $z < 2.5$ are non-hubs. The two dimensional space representation is divided into seven regions, four of them
for non-hub nodes:
(R1) {\em ultra-peripheral nodes}, i.e., nodes with few connections which belong, in turn, to a unique community, 
(R2) {\em peripheral nodes}, which are nodes with few links outside their community,
(R3) {\em non-hub connector nodes}, i.e., nodes with several connections to other communities, and
(R4) {\em non-hub kinless nodes}, with their links homogeneously distributed among all communities.
The other three regions divide the types of hubs into:
(R5) {\em provincial hubs}, i.e., hubs with a large number of their links inside their community,
(R6) {\em connector hubs}, which distribute around $50\%$ of their links in several communities
and (R7) {\em kinless hubs}, whose links are homogeneously distributed among all communities.

\begin{figure}[!t]
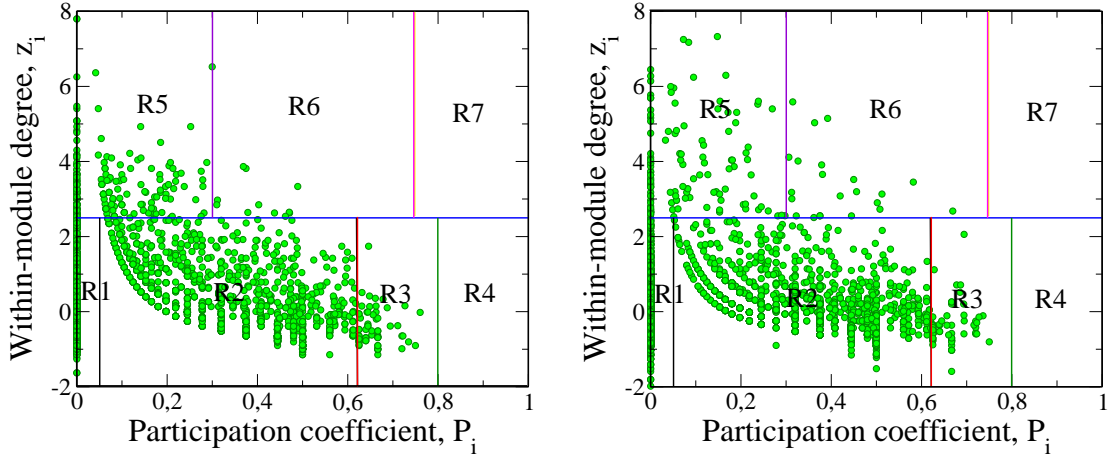

\vskip -0cm
\centerline{
\includegraphics[width=70mm]{fig04a.eps}
\hskip0.5cm
\includegraphics[width=70mm]{fig04b.eps}
}
\vskip0.5cm
\caption{Position of nodes in two dimensional space ($P_i$, $z_i$) for the similarity network (left) and the collaboration network (right). Seven divisions of the two dimensional space used to classify nodes roles are shown explicitly.}
\label{fig:f04}
\end{figure}

In our particular case, we use this classification (after ensuring that it works correctly in our network) in order to identify the central nodes of each community, i.e., the most influencing artists within a particular musical genre, and also those artists who, due to his/her versatility, link two or more musical genres.

\begin{figure}[!b]
\vskip -0cm
\includegraphics[width=140mm]{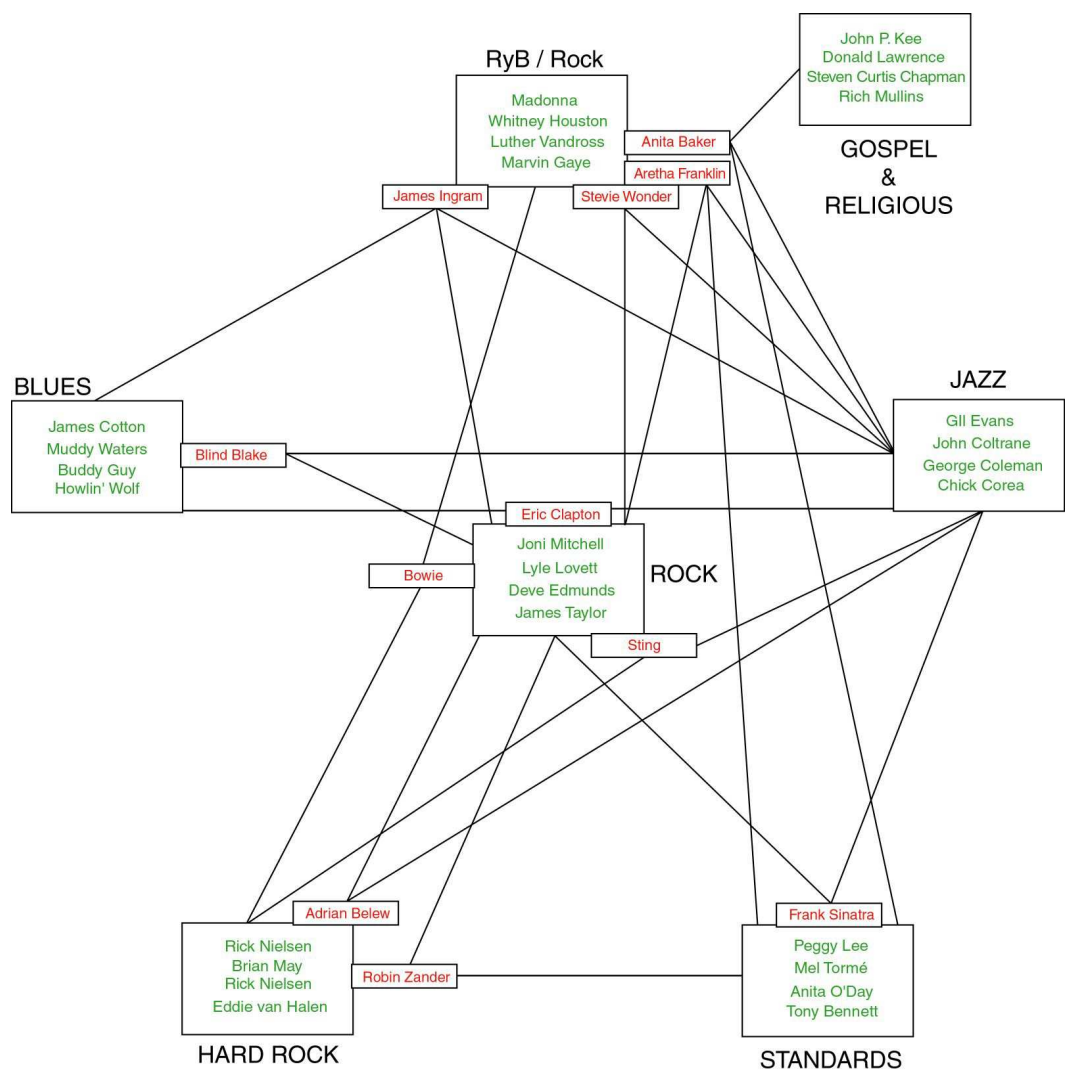}
\caption{
Cartographic representation of the similarity communities. Due to space limits, only the seven largest communities have been plot.
Provincial Hubs (R5, green) and connector hubs (R6,red) have been indicated, in order to show leading artists inside each community and also those artists that act as bridges between musical genres. 
}
\label{fig:f05}
\end{figure}

Figure \ref{fig:f04} shows the position of nodes in the two dimensional space ($P_i$,$z_i$) for both networks.
Provincial hubs of the similarity network (R5) are references in their musical genres. In this category, we find artists such as {\em Elvis Presley, Elton John, Bruce Springsteen, The Rolling Stones, Whitney Houston,
Madonna, Joe Satriani, Axl Rose, John Coltrane or Gil Evans}. On the other hand, there exist artists who are references in their communities but they also stood out for having performed in two or more genres. These artists belong to the R6 category (connector hubs) and we find names as  {\em Stevie Wonder, Eric Clapton, Aretha Franklin, Anita Baker, James Ingram, Sting, David Bowie, Frank Sinatra, Vangelis, Blind Blake, Robin Zander or Adrian Belew}.

\begin{figure}[!t]
\vskip -0cm
\includegraphics[width=140mm]{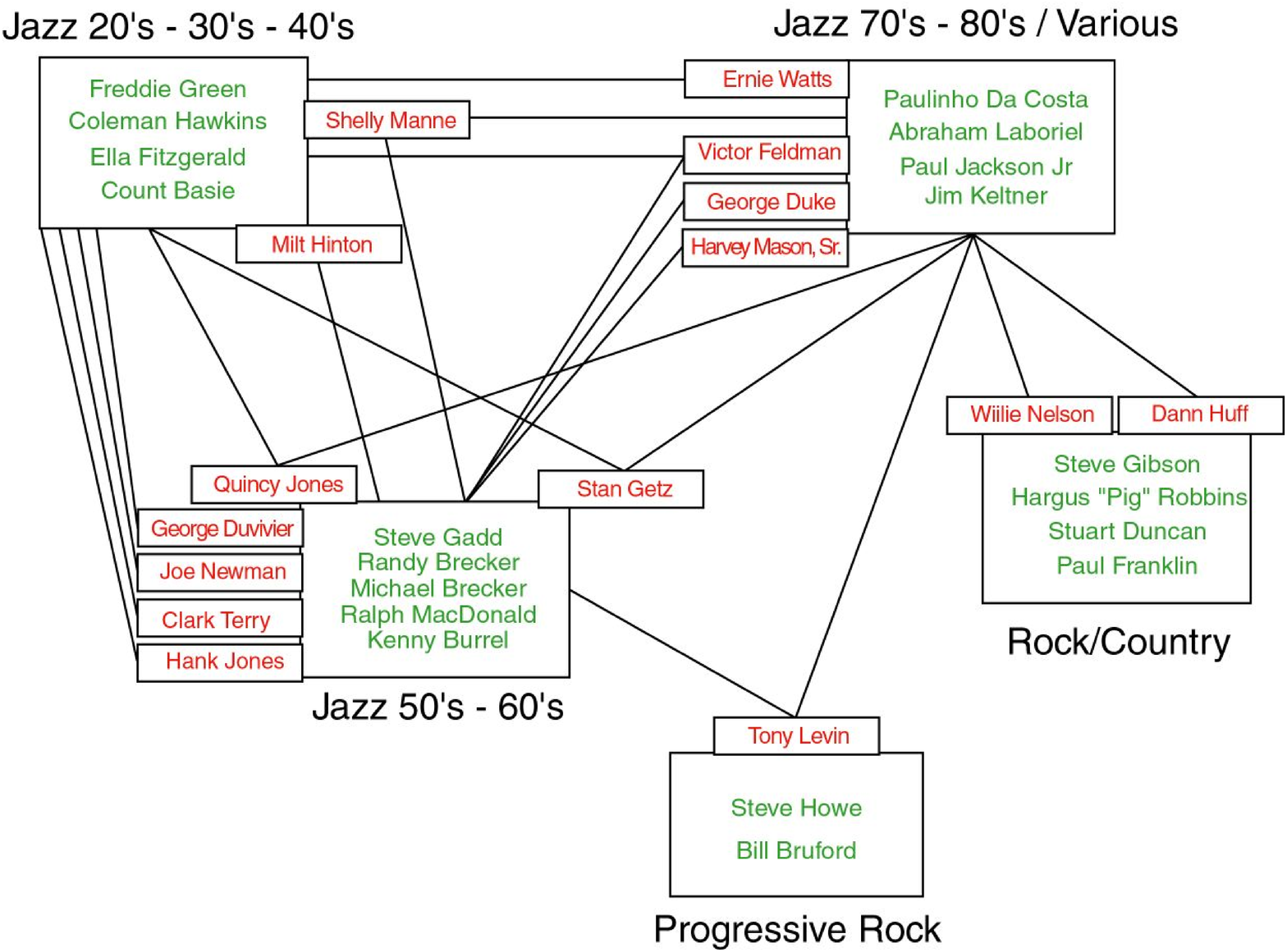}
\caption{
Cartographic representation of the collaboration clusters. Due to space limits, only the five largest communities have been plot.
Provincial Hubs (R5, green) and connector hubs (R6,red) have been indicated, in order to show leading artists inside each community and also those artists that act as bridges between collaboration clusters. 
}
\label{fig:f06}
\end{figure}

In Fig. \ref{fig:f05} and Fig. \ref{fig:f06} we plot a cartographic representation of the seven largest communities within the 
similarity network (Fig. \ref{fig:f05}) and the five largest ones in the collaboration network (Fig. \ref{fig:f06}), where provincial (R5) and
connector (R6) hubs have been explicitly indicated (the rest have been omitted in order to ease the reading. This representation allows
us to identify not only the artists who are references of each musical genre or collaboration clique but also those who act as bridges between communities. 

 As an example, within the {\em Rock} community we can observe how
{\em Eric Clapton} is a connector hub that links the {\em Rock} genre with the {\em Blues} and {\em Jazz} communities. Therefore, Eric Clapton is an internal
connector of the {\em Rock} community. Other kind of connector hub is {\em Blind Blake}, who belongs to the {\em Blues} cluster. This artist is an external 
connector of the {\em Rock} community, since it is one of the bridges between the {\em Blues} and {\em Rock} communities. 
This type of representation provides an objective mechanism for classifying the function of leader artists inside their musical communities by using topological properties of the network and 
furthermore to quantify connections between different musical genres.

\section{Conclusions}

We have shown that the identification of community structures within music networks is a useful tool in order to evaluate the existence
of musical cliques and to identify the role of leading artists inside each community. 
In the case of music similarity networks we have observed that the detected communities are mainly related to musical genres, while the collaboration network presents communities related to artists generations, geographical constraints, genre affinity or music bands. In the collaboration network, for example, jazz players are the most active artists and give rise to the appearance
of large communities related to different generations. 
Finally, we have studied a method to identify the leading artists of each community
and the internal/external connector hubs, who act as bridges between different musical genres.
The information obtained from the community analysis could be a useful tool not only to evaluate the role or relevance of a given artist but to improve the performance of music recommendation systems \cite{sar01,can06,zan07}.  

\section*{Acknowledgements}

P.C. and J.M.B. acknowledge Pablo de Miguel  and T.T. and P.B acknowledge Ariel Chernomoretz for fruitful and encouraging discussions.
This work has been partially supported by project Pharos IST-2006-045035 and the ICREA program. 


\end{document}